\documentclass[12pt, a4paper]{article}
\usepackage{geometry}
\usepackage{amsmath}
\usepackage{graphicx,psfrag,epsf}
\usepackage{enumerate}
\usepackage[hyphens]{url}
\usepackage{setspace}
\usepackage[svgnames]{xcolor}
\usepackage{harvard}
\usepackage{xcolor}
\newcommand{\blind}{1}
\date{\vspace{-5ex}}
\usepackage[utf8]{inputenc}
\usepackage[textsize=small]{todonotes}
\usepackage[normalem]{ulem}
\usepackage{multirow}
\usepackage{multicol}
\usepackage{amsmath}
\usepackage{graphicx,psfrag,epsf}
\usepackage{enumerate}
\usepackage[hyphens]{url}
\usepackage{setspace}
\usepackage[section]{placeins}
\usepackage[utf8]{inputenc}
\usepackage{dcolumn}
\newcolumntype{d}[1]{D{.}{.}{#1}}
\usepackage{booktabs}
\usepackage{siunitx}
\usepackage{multirow}
\usepackage{makecell}
\usepackage{lscape}
\usepackage{ragged2e}
\usepackage{graphicx}
\usepackage[demo]{rotating}
\usepackage{listings}
\usepackage{setspace}
\usepackage{titlesec}
\titlelabel{\thetitle.\quad}
\usepackage[format=hang]{caption}

\usepackage [english]{babel}
\usepackage [autostyle, english = american]{csquotes}
\MakeOuterQuote{"}

\begin{document}

\if1\blind
{
  \title{\bf Generating contingency tables with fixed marginal probabilities and dependence structures described by loglinear models}
  \author{Ceejay Hammond\thanks{Office for National Statistics, London, SW1P 4DF, UK and University of Southampton, Highfield, Southampton, SO17 1BJ, UK. E-mail: cah1g17@soton.ac.uk},
  Peter G.M. van der Heijden\thanks{University of Southampton, Highfield, Southampton, SO17 1BJ and Utrecht University, Padualaan 14, 3584 CH Utrecht, The Netherlands, UK. E-mail: {p.g.m.vanderheijden@uu.nl}} and
Paul A. Smith\thanks{University of Southampton, Highfield, Southampton, SO17 1BJ, UK. E-mail: {p.a.smith@soton.ac.uk}}}
\maketitle
}\fi

\if0\blind
{
\begin{center}
    {\LARGE\bf Generating contingency tables with fixed marginal probabilities and dependence structures described by loglinear models}\\
\hfill \break 
\end{center}
} \fi

\linespread{2}

\newpage
\abstract
\label{abstract}
We present a method to generate contingency tables that follow loglinear models with prescribed marginal probabilities and dependence structures.  We make use of (loglinear) Poisson regression, where the dependence structures, described using odds ratios, are implemented using an offset term. Other statistical models related to loglinear models that fall into the scope of this paper, such as the logistic regression model, the latent class model and the extended Rasch are discussed as well.
We apply this methodology to carry out simulation studies in the context of population size estimation using dual system and triple system estimators, popular in official statistics. These estimators use contingency tables that summarise the counts of elements enumerated or captured within lists that are linked. The simulation is used to investigate these estimators in the situation that the model assumptions are fulfilled, and the situation that the model assumptions are violated. \newline


\newpage
\section{Introduction}
\label{introduction}

In simulation studies that make use of contingency tables, samples are drawn from a population with properties that are well understood. In this paper a method is proposed for the generation of contingency tables with fixed marginal probabilities and prescribed dependence structures that are defined in terms of odds ratios.
An odds ratio describes the strength of the association for a two-way contingency table. For multi-way tables conditional odds ratios play the same role for partial associations between variables (Agresti, 2013)\nocite{AgrestiA2013}. A simple example of such a population would be a two-way contingency table with marginal probabilities of 0.7 and 0.8 and the dependence represented by an odds ratio of 2. 

Loglinear models have interaction parameters that can be directly understood in terms of odds ratios. Therefore the proposed method is built around the use of loglinear models.
Models with two and three variables are explored, where each variable is dichotomous.
However, the methodology presented can generate populations using loglinear models of any size, in terms of both the number of variables and the number of levels of the variables. Also, we can generate populations following logistic regression models having only categorical explanatory variables, and latent variable models that can be formulated as loglinear models, such as the latent class model (Haberman, 1988)\nocite{Haberman1988} and the extended Rasch model (Hessen, 2011)\nocite{HessenDJ2011}.

There are a few ways to fit loglinear models. It suits our purposes to define loglinear models in terms of Poisson regression models with a log link (Nelder and Wedderburn, 1972)\nocite{Nelder1972}. Such models can be fitted with, for example, iteratively reweighted least squares (IRLS) (Green, 1984)\nocite{Green1984}.
The models we use require the marginal probabilities and odds ratios as input parameters. As a first step, the given marginal probabilities are used to generate the probabilities in a loglinear model for which complete independence holds (Good, 1963 and Bishop et al., 1975, p.345) \nocite{BishopYMMFienbergSEHollandPW1975}\nocite{good1963maximum}. The prespecified interaction structure is imposed by using an offset, which is an additional explanatory variable with a prespecified parameter value that is fixed and not estimated.

When each margin of a contingency table has only two categories, the contingency table is a summary of the observations of correlated binary variates. Previously, Lee (1993)\nocite{Lee1993} proposed  a linear programming method for this problem. In comparison to his method, our method is simpler. Also, we present our method in the context of various applications of the loglinear model where conditional odds ratios can also be specified, and of latent variable models. The work of Gange (1995)\nocite{gange1995generating} is related to our work in the sense that he uses the iterative proportional fitting algorithm, an algorithm that was very popular in the early ages of loglinear modelling; yet he does not make a link to loglinear models and mentions odds rations only in passing.
Choi (2008)\nocite{choi2008simple} also presented a method based on loglinear models, where the association is formulated in terms of the Pearson chi-square coefficient, and iterative proportional fitting is used in the generation of probabilities. Thus the way in which the association is specified by Choi is different from our method, that is more closely related to the way in which the loglinear model is defined.
Similar to our work, Kateri  (2014, section 2.2.5, p. 43)\nocite{kateri2014contingency} described that local odds ratios, row and column marginal distribution uniquely determine the table of joint probabilities, which also holds for any other minimal set of odds ratios. Our work further develops this and demonstrates a simple approach to generating data from this information.
Geenens (2020)\nocite{geenens2020copula} sets out a similar approach into a framework where the dependence is measured using copulas. He explains that the bivariate distributions are uniquely determined by a 'margin-free' association parameter (and that multiple different kinds of parameter are possible, but with one-to-one relations between them.

There is also much related but different work on generating correlated binary variates, see, for example, Emrich and Piedmonte (1991)\nocite{emrich1991method}, Lee (1993)\nocite{Lee1993}, Gange (1995)\nocite{gange1995generating}, Park, Park and Shin (1996)\nocite{park1996simple}, Lee (1997)\nocite{lee1997some}, Lunn and Davies (1998)\nocite{lunn1998note}, Al Osh and Lee (2001)\nocite{al2001simple}, Demirtas (2006)\nocite{demirtas2006method} and Jiang, Song, Hou and Zhao (2021)\nocite{jiang2021set}. There is also some (though rather less) work on generating correlated multinomial or ordinal variables which can be used when the margins have more than two categories, for example Gange (1995)\nocite{gange1995generating}, Choi (2008)\nocite{choi2008simple}, Amatya and Demirtas (2015)\nocite{amatya2015multiord} and Touloumis (2016)\nocite{touloumis2016simulating}. These approaches mostly focus on generating population probabilities in order to assess the performance of generalised estimating equation (GEE) models. In passing, some of these papers mention that their methodology can also be used with odds ratios instead of correlations, but they do not provide a detailed description of how this works. Geenens (2020)\nocite{geenens2020copula} extends the connection of copulas and margin-free association measures to general discrete distributions. Combined with marginal information these generate distributions and Geenens demonstrates their existence and uniqueness. Our work is different in that we focus on generating data from loglinear models where the dependence is specified using odds ratios rather than correlations, and describe how this leads to a simple implementation using standard functions for fitting loglinear models.

The methodology we propose is illustrated with an application to population size estimation, as has been used in population censuses. This follows on from the work of Brown, Abbott and Diamond (2006), \nocite{BrownJAbbottODiamondI2006} Baffour, Brown and Smith (2013)\nocite{Baffour2013} and Gerritse, Bakker and van der Heijden (2015) who used a fixed odds ratio
to investigate the impacts of dependence on population size estimation.
In this application, lists where individuals within the target population are enumerated are used to estimate the population sizes. These lists are subject to incompleteness (undercoverage error), and multiple system estimation has been proposed to adjust for this undercoverage error. Some assumptions about the dependence of the lists must be made (though these may be quite weak where there are many lists), so it is important to test these methods 
under prespecified conditions.
The proposed methodology and its application allow investigations into the impacts of given dependence structures within contingency tables to be carried out, varying the sizes of odds ratios, population sizes and response probabilities, and under different models. 

The structure of the paper is as follows. 
In Section \ref{methodology} we outline the methodology around log linear models and model fitting. From this we develop the proposed method for including prespecified interaction parameters in section \ref{model fitting}. 
In Section \ref{appln} we present an application where the proposed method is used to investigate the properties of population size estimation under dependence.
A simulation study is presented for two-way and three-way contingency tables, alongside steps for users to implement the methodology. Section \ref{discussion} makes an assessment of the advantages of our approach compared with existing methods for correlated binary or categorical data.

\section{Methodology}
\label{methodology}

\subsection{Loglinear models}
\label{Section 2.1}

\subsubsection{Models for two-way tables}
\label{m2way}
The simplest situation that we consider is the loglinear model with two dichotomous variables $A$ and $B$. Let $A$ be indexed by $i = 0,1$ and $B$ by $j = 0,1$. We denote the expected count for cell $(i,j)$ of the two-way contingency table of variables $A$ and $B$ by $\mu_{ij}$. We denote the loglinear parameters by $\lambda$. Then the saturated loglinear model for two lists is:

\begin{equation}
\label{sat2}
\log \mu_{ij} = \lambda + \lambda_i^A + \lambda_j^B +\lambda_{ij}^{AB},
\end{equation}

\noindent where $\lambda$ is the intercept term, $\lambda_i^A$ and $\lambda_j^B$ are the respective parameters for variables A and B, and $\lambda_{ij}^{AB}$ is the two factor interaction parameter. The parameters are identified by setting the parameter equal to 0 when an index is 0, i.e. $\lambda_0^A = \lambda_0^B = \lambda_{00}^{AB} = \lambda_{01}^{AB} = \lambda_{10}^{AB} = 0.$ Thus the model has four parameters and four observed counts, and hence is called saturated. The loglinear interaction parameter $\lambda_{11}^{AB}$ is closely related to odds ratio $\theta^{AB}$ and the expected counts by

\begin{equation}
\label{OR}
\theta^{AB} = \mbox{exp}(\lambda_{11}^{AB}) = \frac{(\mu_{11}/\mu_{01})}{(\mu_{10}/\mu_{00})} = 
\frac{\mu_{11}\mu_{00}}{\mu_{01}\mu_{10}}.
\end{equation}

\noindent When $\lambda_{ij}^{AB} = 0$, the odds ratio $\theta^{AB} = 1$, and the independence model holds. 

The proposed methodology can be extended in a straightforward way to variables with more than two levels (compare Kateri, 2014, section 2)\nocite{kateri2014contingency}. Assume a $3 \times 3$ table with levels $i, j = 1,2,3$, with level 3 being the reference category. The model is described in equation (\ref{sat2}) with $i,j$ now having different levels than in the $2 \times 2$ context. If we identify the 9 interaction parameters by setting these parameters equal to 0 when $(i,j = 3)$, then there are four interaction parameters to be estimated, namely $\lambda^{AB}_{11}, \lambda^{AB}_{12}, \lambda^{AB}_{21}$ and $\lambda^{AB}_{22}$. Thus there are four odds ratios, $\theta^{AB}_{11}, \theta^{AB}_{12}, \theta^{AB}_{21}$ and $\theta^{AB}_{22}$, where for each cells in level $(i,j = 3)$ is involved. For example, the odds ratio

\begin{equation}
\label{modeldiag}
\theta^{AB}_{12} = \exp\left(\lambda^{AB}_{12}\right) = \frac{(\mu_{12}/\mu_{13})}{(\mu_{32}/\mu_{33})} =
\frac{\mu_{12}\mu_{33}}{\mu_{13}\mu_{32}}.
\end{equation}

\subsubsection{Models for three-way tables}

\noindent Consider the loglinear model for three dichotomous variables, where the third variable is denoted as $C$. Here the saturated model is

\begin{equation}
\label{sat3}
\log \mu_{ijk} = \lambda + \lambda_i^A + \lambda_j^B + \lambda_k^C +\lambda_{ij}^{AB}+\lambda_{ik}^{AC}+\lambda_{jk}^{BC}+\lambda_{ijk}^{ABC},
\end{equation}

\noindent where $\lambda$ is the intercept term, $\lambda_i^A$ $\lambda_j^B$ and $\lambda_j^C$ are respectively the parameters for variables A, B and C, $\lambda_{ij}^{AB}$, $\lambda_{ik}^{AC}$, and $\lambda_{jk}^{BC}$ are the two factor interaction parameters and $\lambda_{ijk}^{ABC}$ is the three factor interaction parameter. In order to identify the parameters, similar identifying restrictions hold as for the two-variable case.

When three dichotomous variables are present, the odds ratio for each pairwise relationship is conditional on the outcome of the variable not included in the pairwise relationship. E.g., when there is a relationship between variables A and B, the odds ratio is conditional on variable C:

\begin{equation}
\label{model 3}
\theta^{AB|C=0} = \mbox{exp}(\lambda_{11}^{AB}) = \frac{\mu_{110}\mu_{000}}{\mu_{100}\mu_{010}},
\end{equation}

\begin{equation}
\label{model 4}
\theta^{AB|C=1} = \mbox{exp}(\lambda_{11}^{AB} + \lambda_{111}^{ABC}) = \frac{\mu_{111}\mu_{001}}{\mu_{101}\mu_{011}}.
\end{equation}

\subsection{Latent variable models}

Loglinear models are also used in the context of latent variable models. The Rasch model, popular in the field of item response theory in psychometrics and in the field of population size estimation, assumes a single \textit{continuous} latent variable that explains the dependence between  observed dichotomous variables, i.e. the observed variables are conditionally independent given the latent variable. A version of the popular Rasch model (Rasch, 1980, originally published in 1960)\nocite{RaschG1980} is the so-called extended Rasch model (Hessen, 2011), that can be written for three observed variables as model (\ref{sat3}) with all two-factor interactions identical, i.e. $\lambda_{11}^{AB} = \lambda_{11}^{AC} = \lambda_{11}^{BC}$. In psychometric testing the variables $A, B$ and $C$ stand for item responses, and in the Rasch model individuals are ordered by the  difficulty of the items and items are ordered by the aptitude of the individuals. For example, an individual may master item $A$ but neither $B$ nor the most difficult item $C$. Another individual masters $A$ and $B$ and thus has a higher aptitude than the first individual. Thus the model allows individuals to have  heterogeneous probabilities to master items. 

The latent class model assumes the existence of a categorical latent variable $X$ with levels $x$, and given the level $x$ the observed variables are independent. Let's assume that there are four observed variables $A, B, C$ and $D$ (having levels indexed by $l$) and a dichotomous variable $X$. Written in terms of (conditional) probabilities we have,

\begin{equation}\label{LCAprob}
    \pi_{ijkl}^{ABCD} = \Sigma_x \pi_{ijklx}^{ABCDX} = \Sigma_x \pi_x^X\pi_{i|x}^{AX}\pi_{j|x}^{BX}\pi_{j|x}^{CX}\pi_{l|x}^{DX}.
\end{equation}

\noindent This model was originally proposed by Lazersfeld and Henry (1968)\nocite{Lazersfeld1968}. Haberman (1988) \nocite{Haberman1988} showed that the model can also be written as a loglinear model:

\begin{equation}
\label{LCA4}
\log \mu_{ijklx}^{ABCDX} = \lambda + \lambda_i^A + \lambda_j^B + \lambda_k^C + \lambda_l^D+\lambda_{ix}^{AX}+\lambda_{jx}^{BX}+\lambda_{kx}^{CX}+\lambda_{lx}^{DX},
\end{equation}

\noindent with $\mu_{ijkl}^{ABCD} = \Sigma_x \mu_{ijklx}^{ABCDX}$. For the purpose of this paper, writing the latent class model as in equation (\ref{LCA4}) may not be that advantageous to generate a population, as equation (\ref{LCAprob}) can be used for this purpose in a straightforward way by specifying the (conditonal) probabilities on the r.h.s. of equation (\ref{LCAprob}), thus generating population probabilities for the four-way table.

\subsection{Logistic regression}

\noindent Due to the relation between the loglinear model and the logistic regression model with categorical explanatory variables (Bishop \textit{et al}., 1975, section 2.2.4)\nocite{BishopYMMFienbergSEHollandPW1975}, the method proposed in this paper can also be used for this subset of logistic regression models. We show the relation between the models for a simple example. Consider model (\ref{sat3}). Let the variable $C$ with levels $k = 0,1$ be the response variable and variables $A$ and $B$ be categorical variables. Then the logistic regression model is

\begin{equation}
    \log \frac{\mu_{ij1}}{\mu_{ij0}} = \log \mu_{ij1} - \log \mu_{ij0} = \lambda_k^C +\lambda_{ik}^{AC}+\lambda_{jk}^{BC}+\lambda_{ijk}^{ABC},
\end{equation}

\noindent where the $\lambda$-parameters $\lambda_k^C, \lambda_{ik}^{AC}, \lambda_{jk}^{BC}$ and $\lambda_{ijk}^{ABC}$ can be replaced by $\beta$-parameters (not involving $k$ and $C$) $\beta, \beta_{i}^{A}, \beta_{j}^{B}$ and $\beta_{ij}^{AB}$ respectively.
The loglinear model $[AB][C]$ corresponds with the intercept only model in logistic regression as $\lambda_{ik}^{AC} = \lambda_{jk}^{BC} = \lambda_{ijk}^{ABC} = 0$ implies $\beta_{i}^{A} = \beta_{j}^{B} = \beta_{ij}^{AB}$ = 0.

Multinomial and ordinal response models can also be dealt with using this approach if they have equivalent loglinear models. Examples include the baseline category-logit model and the continuation-ratio logit model, see Agresti (2013).\nocite{AgrestiA2013}

\subsection{The proposed method}
\label{model fitting}

Loglinear models are usually estimated using the maximum likelihood method. An important property of maximum likelihood estimates of  loglinear models is that, for the parameters fitted, the corresponding margins of the observed counts equal the margins of the fitted counts (Bishop \textit{et al}., 1975, p.\ 69) \nocite{BishopYMMFienbergSEHollandPW1975}.
These fitted counts are the sufficient statistics. We denote maximum likelihood estimates by adding a hat to parameters and expected counts. 

We focus here on the estimation of hierarchical loglinear models, starting with the two-way table and extending to three-way tables later on. In hierarchical loglinear models, when higher-order parameters are included, the lower order parameters are included as well, for example, if $\lambda_{ij}^{AB}$ is included, the lower order parameters $\lambda_{i}^{A}$ and $\lambda_{j}^{B}$ are included too. The use of hierarchical loglinear models allows for simple and easier interpretation of the model parameters.

\subsubsection{Model fitting for two-way tables}
\label{Section 2.2.1}

For a two-way contingency table with observed counts $n_{ij}$, when we fit model (\ref{sat2}), the fitted values are equal to the observed counts, 
i.e. $\hat{\mu}_{ij} = n_{ij}$. This model is denoted by the variables constituting the highest fitted margins, i.e. [AB]. The margins $n_{ij}$ are the sufficient statistic for the saturated model.  For the independence model, denoted by [A][B], the margins of variable $A$ and variable $B$ are fitted, i.e. 
$\hat{\mu}_{ij} = \hat{\mu}_{i+}\hat{\mu}_{+j}/\hat{\mu}_{++} = n_{i+}n_{+j}/n_{++}$, where a "+" denotes that the sum is taken over the corresponding index.

Although the two most relevant models, the saturated model and the independence model, can be fitted in a straightforward way, we fit the model here through loglinear Poisson
regression, as this suits our purposes later on. Assume that  the four elements $n_{ij}$ are counts derived from four Poisson distributions. Then a likelihood can be set up for $\mu_{ij}$ and maximized over the parameters. If we collect the elements $m_{ij}$ into a column vector, then  model (\ref{sat2}) can be written as,

\begin{equation}
\label{matrixeq}
  \left[ {\begin{array}{c}
    \log(\mu_{11}) \\
    \log(\mu_{10}) \\
    \log(\mu_{01}) \\
    \log(\mu_{00}) \\
  \end{array} } \right]
  =
  \left[ {\begin{array}{cccc}
    1 & 1 & 1 & 1 \\
    1 & 1 & 0 & 0 \\
    1 & 0 & 1 & 0 \\
    1 & 0 & 0 & 0 \\
  \end{array} } \right]
   \left[ {\begin{array}{c}
    \lambda\\
    \lambda_{1}^{A}\\
    \lambda_{1}^{B}\\
    \lambda_{11}^{AB}\\
  \end{array} } \right],
\end{equation}

\noindent where the matrix with 0's and 1's is the design matrix formulating the loglinear model. Maximizing the likelihood gives maximum likelihood estimates $\hat{\mu}_{ij}$ and parameter estimates $\hat{\lambda}, \hat{\lambda}_1^A, \hat{\lambda}_1^B$ and $\hat{\lambda}_{11}^{AB}$. In fitting the independence model, the last column of the design matrix, $(1, 0, 0, 0)^T$ and the parameter $\lambda_{11}^{AB}$ are left out of the equation. In software packages this can be accomplished in routines for Poisson regression, where the four observed counts $n_{ij}$ are provided as the $Y$-variable, the model is represented by the design matrix in equation (\ref{matrixeq}), and the output consists of four fitted values $\hat{\mu}_{ij}$ and the parameter estimates.

\subsubsection{Generating two-way tables with given properties}
We now develop  
a model for the two-way contingency table with prespecified marginal probabilities and a prespecified interaction parameter. 
For the interaction parameter we want to make use of the odds ratio $\theta^{AB}$ defined in equation (\ref{OR}).
This can be accomplished by using an offset in Poisson regression. An offset is a column in the design matrix for which no parameter is estimated, i.e. the parameter is fixed to 1.
The odds ratio is specified to give the desired level of dependence between lists. If we denote the prespecified odds ratio by $\tilde{\theta}^{AB}$, then the desired loglinear interaction parameter is $\tilde{\lambda}_{11}^{AB} = \log \tilde{\theta}^{AB}$. Thus the population that we want to generate has to follow the following model:

\begin{equation}
\label{sat2tilde}
\log \mu_{ij} = \lambda + \lambda_i^A + \lambda_j^B + \tilde{\lambda}_{ij}^{AB}.
\end{equation}

\noindent The following equation, an adjusted version of equation (\ref{matrixeq}), illustrates how this is implemented in Poisson regression:

\begin{equation}
    \label{matrixeq2offset}
  \left[ {\begin{array}{c}
    \log(\mu_{11}) \\
    \log(\mu_{10}) \\
    \log(\mu_{01}) \\
    \log(\mu_{00}) \\
  \end{array} } \right]
  =
  \left[ {\begin{array}{cccc}
    1 & 1 & 1 & \log \tilde{\theta}^{AB} \\
    1 & 1 & 0 & 0 \\
    1 & 0 & 1 & 0 \\
    1 & 0 & 0 & 0 \\
  \end{array} } \right]
   \left[ {\begin{array}{c}
    \lambda\\
    \lambda_{1}^{A}\\
    \lambda_{1}^{B}\\
    1\\
  \end{array} } \right].
 \end{equation}

\noindent Notice that the parameter $\lambda^{AB}_{11}$ from (\ref{matrixeq}) is set to 1, i.e. it is not estimated, and this value 1 is multiplied with the prespecified $\log \tilde{\theta}^{AB}$ in the design matrix. In software packages for Poisson regression this can be accomplished as follows:
\begin{itemize}
\item{specify an offset vector which is $\log\tilde{\theta}^{AB}$ when $(i,j) = (1,1)$ and 0 otherwise.}
\item{use an independence model as the design matrix, as we only need estimates for $\lambda, \lambda_i^A$ and $\lambda_j^B$.}
\item{in order to fit a model that leads to estimated counts that follow the prespecified marginal probabilities, calculate joint probabilities for which these marginal probabilities hold, and use these joint probabilities, multiplied with some constant, as the counts $n_{ij}$ to be analysed with model (\ref{sat2tilde}).}
\item{then the estimates of expected frequencies can be used to derive the joint probabilities with the prespecified marginal probabilities and the prespecified odds ratio.}
\end{itemize}

\noindent As an example, assume we want to generate population probabilities that have the prespecified marginal probabilities 0.7 and 0.8 and the prespecified odds ratio 2. Then the probabilities to be analysed are 0.7 * 0.8 = 0.56, 0.7 * 0.2 = 0.14, 0.3 * 0.8 = 0.24 and 0.3 * 0.2 = 0.06, and multiply these probabilities with some number, such as 1000, to get counts. 
Thus 560, 140, 240 and 60 are the values of the dependent variable in Poisson regression. Specify $\log (2)$ as the offset and analyse the counts with the loglinear independence model and the offset. This yields the fitted values 584.8, 115.2, 215.2, 84.8  (rounded to 1 decimal place),
and by dividing them by 1000 we have the  four probabilities, with the prespecified marginal probabilities 0.7 and 0.8 and the prespecified odds ratio 2. The R-code we used can be found in the online Appendix.

We now prove that for model (\ref{sat2tilde}) the prespecified margins of the observed table are equal to those in the fitted table, irrespective of the offset. We follow Fienberg (1980, Appendix II)\nocite{FienbergS1980}. If we assume that the counts are observations from independent Poisson distributions, then the likelihood function is proportional to

\begin{equation}
    \prod_{ij} \mu_{ij}^{n_{ij}} \exp (-\mu_{ij}),
\end{equation}
and the kernel of the loglikelihood is,
\begin{equation}
    \sum_{ij} n_{ij} \log \mu_{ij} - \sum_{ij} \mu_{ij} = 
    \sum_{ij} n_{ij}\left(\lambda + \lambda_i^A + \lambda_j^B + \tilde{\lambda}_{ij}^{AB}\right) - 
    \sum_{ij} \mu_{ij}.
\end{equation}
At the maximum of the likelihood the derivate of the loglikelihood over the parameters is 0. So, for example, taking the derivative over $\lambda_i^A$ gives $n_{i+} - \hat{\mu}_{i+} = 0$ and this gives $n_{i+} = \hat{\mu}_{i+}$. Similarly for $n_{+j} = \hat{\mu}_{+j}$. This completes the proof. It follows that including an offset for the interaction does not affect the property of ordinary loglinear model that for the fitted parameters the corresponding margins of the fitted values are equal to those of the observed values.

The model for the $3\times3$ table in section \ref{m2way} has the following implementation:
\begin{equation}
    \label{matrixdiagoffset}
  \left[ {\begin{array}{c}
    \log(\mu_{11}) \\
    \log(\mu_{12}) \\
    \log(\mu_{13}) \\
    \log(\mu_{21}) \\
    \log(\mu_{22}) \\
    \log(\mu_{23}) \\
    \log(\mu_{31}) \\
    \log(\mu_{32}) \\
    \log(\mu_{33}) \\
  \end{array} } \right]
  =
  \left[ {\begin{array}{cccccc}
    1 & 1 & 0 & 1 & 0 & \log \tilde{\theta}^{AB}_{11}\\
    1 & 1 & 0 & 0 & 1 & \log \tilde{\theta}^{AB}_{12}\\
    1 & 1 & 0 & 0 & 0 & 0 \\
    1 & 0 & 1 & 1 & 0 & \log \tilde{\theta}^{AB}_{21}\\
    1 & 0 & 1 & 0 & 1 & \log \tilde{\theta}^{AB}_{22}\\
    1 & 0 & 1 & 0 & 0 & 0 \\
    1 & 0 & 0 & 1 & 0 & 0 \\
    1 & 0 & 0 & 0 & 1 & 0 \\
    1 & 0 & 0 & 0 & 1 & 0 \\
  \end{array} } \right]
   \left[ {\begin{array}{c}
    \lambda\\
    \lambda_{1}^{A}\\
    \lambda_{2}^{A}\\
    \lambda_{1}^{B}\\
    \lambda_{2}^{B}\\
    1\\
  \end{array} } \right].
 \end{equation}

\subsubsection{Model fitting for three-way tables}
\label{Section 2.2.2}

\noindent For the saturated model with three variables, the fitted values are equal to the observed counts
so $\hat{\mu}_{ijk} = n_{ijk}$. As an example of a restricted model, the estimates for the [AB][AC] model can be calculated directly as 

\begin{equation}
\label{AB,AC}
\hat{\mu}_{ijk} = \frac{\hat{\mu}_{ij+}\hat{\mu}_{i+k}}{\hat{\mu}_{i++}} = \frac{n_{ij+}n_{i+k}}{n_{i++}}.
\end{equation}

\noindent However, for the loglinear model [AB][AC][BC] estimates must be calculated indirectly, iteratively (for example, Bishop \textit{et al}., 1975, p.84).\nocite{BishopYMMFienbergSEHollandPW1975}

We now develop (\ref{sat3}) in the form of Poisson regression. There are eight elements $n_{ijk}$, being counts from eight Poisson distributions. The likelihood can be set up for $\mu_{ijk}$ and maximised over the parameters. Model (\ref{sat3}) can be denoted as,

\begin{equation}
    \label{matrixeq3}
  \left[ {\begin{array}{c}
    \log(\mu_{111}) \\
    \log(\mu_{110}) \\
    \log(\mu_{101}) \\
    \log(\mu_{100}) \\
    \log(\mu_{011}) \\
    \log(\mu_{010}) \\
    \log(\mu_{001}) \\
    \log(\mu_{000}) \\
  \end{array} } \right]
  =
  \left[ {\begin{array}{cccccccc}
    1 & 1 & 1 & 1 & 1 & 1 & 1 & 1 \\
    1 & 1 & 1 & 0 & 1 & 0 & 0 & 0 \\
    1 & 1 & 0 & 1 & 0 & 1 & 0 & 0 \\
    1 & 1 & 0 & 0 & 0 & 0 & 0 & 0 \\
    1 & 0 & 1 & 1 & 0 & 0 & 1 & 0 \\
    1 & 0 & 1 & 0 & 0 & 0 & 0 & 0 \\
    1 & 0 & 0 & 1 & 0 & 0 & 0 & 0 \\
    1 & 0 & 0 & 0 & 0 & 0 & 0 & 0 \\
  \end{array} } \right]
   \left[ {\begin{array}{c}
    \lambda\\
    \lambda_{1}^A\\
    \lambda_{1}^B\\
    \lambda_{1}^C\\
    \lambda_{11}^{AB}\\
    \lambda_{11}^{AC}\\
    \lambda_{11}^{BC}\\
    \lambda_{111}^{ABC}\\
  \end{array} } \right].
\end{equation}

\noindent As before, for (\ref{matrixeq3}) the matrix with 0's and 1's is the design matrix formulating the loglinear model. A restricted model can be fitted by dropping columns in equation (\ref{matrixeq3}). For example, in fitting the model where $B$ and $C$ are independent given $A$,  the last two columns of the design matrix and the last two parameters  are left out of the equation.

\subsubsection{Generating three-way tables with given properties}
We now focus on loglinear models with prescribed odds ratios. We start with the model with no three-factor interaction. The model for a three-way contingency table with three prespecified odds ratios $\tilde{\theta}^{AB}, \tilde{\theta}^{AC}$ and $\tilde{\theta}^{BC}$ is denoted as,
\begin{equation}
\label{sat3tilde}
\log \mu_{ij} = \lambda + \lambda_i^A + \lambda_j^B + \lambda_k^C + \log\tilde{\theta}^{AB}+ \log\tilde{\theta}^{AC}+ \log\tilde{\theta}^{BC}.
\end{equation}

\noindent In the adjusted version of (\ref{matrixeq3}) this becomes (using that $\log\tilde{\theta}^{AB} +  \log\tilde{\theta}^{AC} + \log\tilde{\theta}^{BC} = \log\tilde{\theta}^{AB}\tilde{\theta}^{AC}\tilde{\theta}^{BC}$):

\begin{equation}
    \label{matrixeq4offset}
  \left[ {\begin{array}{c}
    \log(\mu_{111}) \\
    \log(\mu_{110}) \\
    \log(\mu_{101}) \\
    \log(\mu_{100}) \\
    \log(\mu_{011}) \\
    \log(\mu_{010}) \\
    \log(\mu_{001}) \\
    \log(\mu_{000}) \\
  \end{array} } \right]
  =
  \left[ {\begin{array}{ccccc}
    1 & 1 & 1 & 1 & \log\tilde{ \theta}^{AB}\tilde{\theta}^{AC}\tilde{ \theta}^{BC} \\
    1 & 1 & 1 & 0 & \log\tilde{ \theta}^{AB} \\
    1 & 1 & 0 & 1 & \log\tilde{ \theta}^{AC} \\
    1 & 1 & 0 & 0 & 0 \\
    1 & 0 & 1 & 1 & \log\tilde{ \theta}^{BC} \\
    1 & 0 & 1 & 0 & 0\\
    1 & 0 & 0 & 1 & 0\\
    1 & 0 & 0 & 0 & 0\\
  \end{array} } \right] 
   \left[ {\begin{array}{c}
    \lambda \\
    \lambda_{1}^A \\
    \lambda_{1}^B \\
    \lambda_{1}^C \\
    1\\
  \end{array} } \right],
\end{equation}

\noindent where the parameter for the offsets has been set to 1. In software packages for Poisson regression this can be accomplished as follows:
\begin{itemize}
\item{specify $\log\tilde{ \theta}^{AB}$, $\log \tilde{\theta}^{AC}$ and $\log\tilde{ \theta}^{BC}$, and use them to generate the offset column of the design matrix as in (\ref{matrixeq4offset}).}
\item{use an independence model in the design matrix, as we only need estimates for $\lambda, \lambda_i^A$, $\lambda_j^B$ and $\lambda_k^C$.} 
\item{in order to fit a model that leads to estimated counts that follow prespecified marginal probabilities, calculate joint probabilities
for which these marginal probabilities hold by multiplying the relevant probabilities. Use these joint probabilities, multiplied by some constant representing a notional population size, as the observed counts corresponding with $n_{ijk}$ to be analysed with model (\ref{matrixeq4offset}).}
\item{then, as before in Section \ref{Section 2.2.1}, the estimates of the expected frequencies can be used to derive the joint probabilities with the prespecified marginal probabilities and the prespecified odds ratios.}
\end{itemize}

\noindent As an example, assume we want to generate population probabilities that have the prespecified marginal probabilities 0.8, 0.7 and 0.9  for variables $A, B$ and $C$ respectively, and with all the prespecified odds ratios $= 2$. 
The probabilities following the independence model are 0.7 * 0.8 * 0.9 = 0.504, 0.7 * 0.8 * 0.1 = 0.056, 0.7 * 0.2 * 0.9 = 0.216, 0.7 * 0.2 * 0.1 = 0.024, 0.3 * 0.8 * 0.9 = 0.126, 0.3 * 0.8 * 0.1 = 0.014, 0.3 * 0.2 * 0.9 = 0.054 and 0.3 * 0.2 * 0.1 = 0.006 and we multiply these probabilities by 1,000 to get counts 504, 56, 216, 24, 126, 14, 54 and 6 which form the values of the dependent variable in Poisson regression. Specify $\log (2)$  as the three offsets and analyse the counts with the loglinear independence model and the offset as in equation (\ref{matrixeq4offset}). This yields the fitted values (rounded to 1 decimal place) 547.3, 39.5, 186.3, 26.9, 99.0, 14.3, 67.4 and 19.4. By dividing these by 1000 we have the eight probabilities with the prespecified marginal probabilities 0.7, 0.8 and 0.9 and the three prespecified odds ratios 2. For example, the marginal probability of being in variable A equals the sum of the fitted values for $A = 1$ divided by 1000, which is 0.8 and the conditional odds $\theta^{BC|A=1} = (547.3 * 26.9)/(39.5 * 186.3) = 2$. The R-code we used can be found in the online Appendix.

We now discuss the model with a prespecified three-factor interaction.
Above we discussed model fitting for three-way tables, with no three-factor interaction. Here we develop the model to include a prespecified three-factor interaction parameter in a loglinear model. The three-factor interaction is defined as the state where two conditional odds ratios are not equal, for example, $\tilde{\theta}^{AB|C=0} \neq \tilde{\theta}^{AB|C=1}$. These conditional odds ratios are related to loglinear parameters by $\tilde{\theta}^{AB|C=0} = \exp \tilde{\lambda}^{AB}_{11}$ and 
$\tilde{\theta}^{AB|C=1} = \exp \left[\tilde{\lambda}^{AB}_{11} + \tilde{\lambda}^{ABC}_{111}\right]$. As similar results hold for $\tilde{\theta}^{AC|B=0}$, $\tilde{\theta}^{AC|B=1}$, $\tilde{\theta}^{BC|A=0}$ and $\tilde{\theta}^{BC|A=1}$ this shows that the three-way interaction is represented by $\exp \tilde{\lambda}^{ABC}_{111}$.
A three-way interaction will be handled as follows.
In Equation (\ref{matrixeq4offset}) we fill in $\log\tilde{\theta}^{AB|C=0}$ in cell (2,5) of the design matrix and $\log\tilde{\theta}^{AB|C=1}$ in cell (1,5) of the design matrix, and similarly for the other conditional odds ratios. 
Thus, the model with prespecified conditional odds ratios can be presented as,

\begin{equation}
    \label{matrixeq5offset}
  \left[ {\begin{array}{c}
    \log(\mu_{111}) \\
    \log(\mu_{110}) \\
    \log(\mu_{101}) \\
    \log(\mu_{100}) \\
    \log(\mu_{011}) \\
    \log(\mu_{010}) \\
    \log(\mu_{001}) \\
    \log(\mu_{000}) \\
  \end{array} } \right]
  =
  \left[ {\begin{array}{ccccc}
    1 & 1 & 1 & 1 & \log\tilde{\theta}^{AB|C=1}\tilde{\theta}^{AC|B=1}\tilde{\theta}^{BC|A=1} \\
    1 & 1 & 1 & 0 & \log\tilde{ \theta}^{AB|C=0} \\
    1 & 1 & 0 & 1 & \log\tilde{ \theta}^{AC|B=0} \\
    1 & 1 & 0 & 0 & 0 \\
    1 & 0 & 1 & 1 & \log\tilde{ \theta}^{BC|A=0} \\
    1 & 0 & 1 & 0 & 0\\
    1 & 0 & 0 & 1 & 0\\
    1 & 0 & 0 & 0 & 0\\
  \end{array} } \right] 
   \left[ {\begin{array}{c}
    \lambda \\
    \lambda_{1}^{A} \\
    \lambda_{1}^{B} \\
    \lambda_{1}^{C} \\
    1\\
  \end{array} } \right].
\end{equation}

\noindent We can trivially write

\begin{equation}
\tilde{\theta}^{AB|C=1} = \tilde{\theta}^{AB|C=0}\frac{\tilde{\theta}^{AB|C=1}}{\tilde{\theta}^{AB|C=0}},
\end{equation}

\noindent the term $\left[\tilde{\theta}^{AB|C=1}/\tilde{\theta}^{AB|C=0}\right]$, which is the difference in odds for $C=1$ and $C=0$, is equal to $\exp \tilde{\lambda}^{ABC}_{111}$. The terms $\tilde{\theta}^{AC|B=1}$ and $\tilde{\theta}^{BC|A=1}$ include this same three-way interaction term, i.e.
\begin{equation}
\tilde{\theta}^{AC|B=1} = \tilde{\theta}^{AC|B=0}\frac{\tilde{\theta}^{AB|C=1}}{\tilde{\theta}^{AB|C=0}},
\end{equation}
\begin{equation}
\tilde{\theta}^{BC|A=1} = \tilde{\theta}^{BC|A=0}\frac{\tilde{\theta}^{AB|C=1}}{\tilde{\theta}^{AB|C=0}}.
\end{equation}
\noindent In fact since the three-factor interaction parameter
$\tilde{\lambda}_{111}^{ABC}$ is the difference between each pair of  conditional odds ratios, it is clear that
\begin{equation}\frac{\theta^{AB|C=1}} {\theta^{AB|C=0}} = \frac{\theta^{AC|B=1}} {\theta^{AC|B=0}} = \frac{\theta^{BC|A=1}} {\theta^{BC|A=0}}.
\label{equalodds}
\end{equation}  
If we specify the denominators of each fraction, then as soon as one numerator is specified then all the fractions are determined. There are therefore only four free odds ratios in equation (\ref{equalodds}), and the three-factor interaction can be specified by providing all the denominators and any one of the numerators. Therefore we can rewrite element (1,5) of the design matrix in (\ref{matrixeq5offset}) as
\begin{multline}
    \log\tilde{\theta}^{AB|C=1}\tilde{\theta}^{AC|B=1}\tilde{\theta}^{BC|A=1} = \log\tilde{\theta}^{AB|C=0}\tilde{\theta}^{AC|B=0}\tilde{\theta}^{BC|A=0}\left[\frac{\tilde{\theta}^{AB|C=1}}{\tilde{\theta}^{AB|C=0}}\right]^3,
\end{multline}
using only four odds ratios.

As an example, assume that we want to create population probabilities with conditional odds ratios $\tilde{\theta}^{AB|C=0} = 2$ and  $\tilde{\theta}^{AB|C=1} = 3$. It follows that the three factor term is $\left[\tilde{\theta}^{AB|C=1}/\tilde{\theta}^{AB|C=0}\right] = 3/2$. Then, let $\tilde{\theta}^{AC|B=1} = 1$, it follows that $\tilde{\theta}^{AC|B=0}\left[\tilde{\theta}^{AB|C=1}/\tilde{\theta}^{AB|C=0}\right] = 1$, so that $\tilde{\theta}^{AC|B=0} = 2/3$. Similarly, if $\tilde{\theta}^{BC|A=1} = 1$, then $\tilde{\theta}^{BC|A=0} = 2/3$. Thus the model equation becomes

\begin{equation}
    \label{matrixeq6offset}
  \left[ {\begin{array}{c}
    \log(\mu_{111}) \\
    \log(\mu_{110}) \\
    \log(\mu_{101}) \\
    \log(\mu_{100}) \\
    \log(\mu_{011}) \\
    \log(\mu_{010}) \\
    \log(\mu_{001}) \\
    \log(\mu_{000}) \\
  \end{array} } \right]
  =
  \left[ {\begin{array}{ccccc}
    1 & 1 & 1 & 1 & \log 3\\
    1 & 1 & 1 & 0 & \log 2 \\
    1 & 1 & 0 & 1 & \log (2/3) \\
    1 & 1 & 0 & 0 & 0 \\
    1 & 0 & 1 & 1 & \log (2/3) \\
    1 & 0 & 1 & 0 & 0\\
    1 & 0 & 0 & 1 & 0\\
    1 & 0 & 0 & 0 & 0\\
  \end{array} } \right] 
   \left[ {\begin{array}{c}
    \lambda \\
    \lambda_{1}^{A} \\
    \lambda_{1}^{B} \\
    \lambda_{1}^{C} \\
    1\\
  \end{array} } \right].
\end{equation}

\section{Application to Population Size Estimation}
\label{appln}

\subsection{Dual System Estimation}

Nowadays dual system estimation methods are commonly used to estimate population totals for domains of interest. The dual system estimator is the simplest capture-recapture model and is used to estimate the part of the population that is missing after having linked individuals in two overlapping lists. Originally this method was implemented to estimate the size and density of wildlife populations, but it has also been developed and used in official statistics, where the dual-system estimator was originally used for the 1950 decennial United States Census to estimate coverage error (Wolter, 1986)\nocite{WolterKM1986}.

The linked data can be presented in a two-way contingency table: we observe the counts of those in both lists, those in list A only, those in list B only, but not those missing from both lists. Thus there are three observed cell counts, and the missing count for those in neither of the lists is to be estimated, which can conveniently be done using the loglinear model framework (Fienberg, 1972)\nocite{FienbergS1972}.

Dual system estimation relies on five key assumptions, outlined by the International Working Group for Disease Monitoring and Forecasting (1995)\nocite{IWGfDMaF1995}:

\begin{enumerate}
    \item The population is closed, i.e. there is no change such as births or deaths.
    \item There is perfect matching between the two lists.
    \item The capture/inclusion probability of elements in at least one of the lists is homogeneous (Seber, 1982;  van der Heijden \textit{et al}., 2012).\nocite{Seber1982} \nocite{VanderHeijdenPGMWhittakerCruyffBakkerVanderVliet2012} 
    \item The probabilities of inclusion in the lists are independent.
    \item There are no erroneous captures in any list (no overcoverage).
\end{enumerate}

One of the key assumptions is that the probabilities of inclusion in the lists are independent, which is easily violated (Gerritse \textit{et al}., 2015)\nocite{GerritseBakkervanderHeijden2015}. However, with the data used for population size estimation, this assumption is impossible to verify and so the design of both lists is very important, to create the conditions where this assumption is likely to hold. A way to reduce the impact of violating this strict assumption is to divide the population by certain characteristics, such as age, sex and some geographical breakdown, so that we need only assume that the lists are independent conditional on these classifiers.

Assumption 4 follows on from assumption 3, which implies that individuals who where included in the first list and those who were not have the same probability of being included in the second list, so that presence in the first list does not affect response in the second list, and the lists are therefore statistically independent (International Working Group for Disease Monitoring and Forecasting, 1995)\nocite{IWGfDMaF1995}.
Therefore, as there are three observed counts, to estimate cell counts $\hat{\mu}_{ij}$ for a two-way contingency table only three parameters can be used and no interaction term can be modelled, and (\ref{sat2}) is adjusted to include no interaction term, i.e. $\lambda_{ij}^{AB} = 0$. Independence implies that the odds ratio (\ref{OR}) is 1, so that the missing count is estimated by $\hat{\mu}_{00} = \hat{\mu}_{10}\hat{\mu}_{01}/\hat{\mu}_{11} = n_{10}n_{01}/n_{11} $.

\subsection{Triple System Estimation}

The triple system estimator uses three linked lists. Counts can be summarised in a three-way contingency table where one count is missing, namely the count for the individuals that are missed by all three lists. By estimating this count and adding it to the seven observed counts, the population size is estimated. Unlike a two-way contingency table, in a three-way contingency table the pairwise interactions between lists can be modelled. 
This means that the homogeneous capture/inclusion probabilities and the independence assumption can be relaxed, while the other assumptions outlined for dual system estimation still hold.

As there are seven cells with observed counts, only loglinear models with up to seven independent parameters can be fitted. This is usually solved by assuming the three-factor interaction term $\hat{\lambda}_{ijk}^{ABC} = 0$. Including two-factor interaction terms in the model enables us to estimate population totals where pairwise relationships are present (Bishop \textit{et al}., 1975)\nocite{BishopYMMFienbergSEHollandPW1975}.
For this paper, we are interested in fitting the maximal model for a three-way contingency table with one unobserved cell, which is an adjusted version of (\ref{sat3}) with no three factor interaction term.\newline

\subsection{How to apply the proposed methodology to population size estimation}

The following outlines how to generate contingency tables with prespecified margins and interactions between lists. Such generated contingency tables can be used in simulations to study the behaviour of estimators for the population size, as the true population size is known.

To generate a contingency table where lists are independent, the total population size and inclusion probabilities for each list are prespecified. The cell counts are then computed by multiplying the marginal inclusion probabilities and the population size. This results in cell counts that follow statistical independence between lists. 

To generate a table with a given dependence structure, an independent loglinear model is then fitted to these cell counts with the inclusion of an offset term,  as presented in (\ref{sat2tilde}) and (\ref{sat3tilde}). Then the fitted counts have the following properties:

\begin{enumerate}
    \item Their sum is equal to the prespecified population size, 
    \item The prespecified odds ratios hold, and
    \item The prespecified marginal response probabilities hold.
\end{enumerate}

In order to investigate the behavior of population size estimators, multinomial samples are drawn from this population. For the study of the dual system estimator, in each sample the cell (0,0) is set to missing and the estimator is applied. For the study of the triple systems estimator, in each sample the cell (0,0,0) is set to missing and the estimator is applied. When $t$ samples are drawn, $t$ population size estimates are obtained that can be compared with the prespecified population size.
\newline

\subsection{Simulation Study}

The aim of the simulation study is to illustrate the methodology. Probabilities are generated using prespecified marginal probabilities and odds ratios. 2000 multinomial samples are drawn with these probabilities and with the total population size  set equal to 1000 or 2000. The prespecified probabilities and odds ratios mirror those used by Brown, Abbott and Diamond (2006)\nocite{BrownJAbbottODiamondI2006} in their investigation of methods to deal with dependence in the coverage estimation for the 2001 population census of England and Wales. The outcomes of the simulations are presented in Table \ref{table:4} and Table \ref{table:6}. The results include the specified values for the total population, the marginal response probabilities, the odds ratio (OR) between the lists, the mean and median of the population size estimates, empirical 95\% confidence intervals (which is the range between the 2.5\%) and 97.5\% quantiles of the estimates, the Relative Bias \%, ($\hat{\mu}$ - $N$)/$N$ and the Coefficient of Variation $\hat{\sigma}$/$N$, where $N$ is the prespecified population size, $\hat{\mu}$ is the mean and $\hat{\sigma}$ is the standard deviation of the population size estimates over the simulations.

\subsubsection{Two-way table}

We set the response probabilities to either, for list A ($\pi_A$) 0.8 and for list B ($\pi_B$) 0.7; or for list A ($\pi_A$) 0.9 and for list B ($\pi_B$) 0.8. The joint probabilities ($\pi_{11}, \pi_{10}, \pi_{01}, \pi_{00}$) are, respectively, (0.56, 0.24, 0.14, 0.06) and (0.72, 0.18, 0.08 and 0.02). Cell counts are obtained by $\mu_{ij}=(N\pi_{ij})$.

\begin{table}[!htb]\footnotesize
\centering
\captionsetup{width=4.9in}
\caption{\small Simulation results for dual systems estimation with inclusion probabilities $\pi_A$ and  $\pi_B$, sizes of odds ratios (OR), the mean and median of the estimates, the 95\% confidence interval of the mean, the relative bias, and coefficient of variation (CV), where the fitted model is [A][B]}
\begin{tabular}{ccccrrrrrr*{1}{S[table-format=2]}}  
 \hline
 N & $\pi_A$ & $\pi_B$ & OR & \multicolumn{1}{c}{Mean} &  \multicolumn{1}{c}{Median} &  \multicolumn{2}{c}{95\%CI} & \multicolumn{1}{c}{rbias\%} & \multicolumn{1}{c}{CV}
 \\
 \hline
 1000 & 0.8 & 0.7 & 1 & 1000.3  & 1000.0 & 979.5,& 1020.6 & 0.027 & 0.011\\
 1000 & 0.9 & 0.8 & 1 & 1000.0  & 1000.0 & 989.8, & 1010.0 &-0.005& 0.005\\
 2000 & 0.8 & 0.7 & 1 & 2000.1  & 2000.2 & 1971.3,& 2029.1 & 0.005& 0.007\\
 2000 & 0.9 & 0.8 & 1 & 2000.1 & 2000.1 & 1985.2,& 2014.6&0.007 & 0.004\\
 \hline
 1000 & 0.8 & 0.7 & 1.1 & 994.5  & 994.5 & 973.9, & 1015.2&-0.545&0.011\\
 1000 & 0.9 & 0.8 & 1.1 & 998.0  & 998.1 & 986.9, & 1008.4&-0.197&0.005\\
 2000 & 0.8 & 0.7 & 1.1 & 1989.9  & 1989.1 & 1960.3,& 2018.0&-0.553&0.007\\
 2000 & 0.9 & 0.8 & 1.1 & 1996.1 & 1996.0& 1980.9,& 2010.2&-0.197&0.004\\
 \hline
 1000 & 0.8 & 0.7 & 2   & 957.8  & 957.8 & 937.5, & 978.9&-4.215&0.011\\
 1000 & 0.9 & 0.8 & 2   & 984.4  & 984.6 & 972.3, & 995.9&-1.560&0.006\\
 2000 & 0.8 & 0.7 & 2   & 1915.6  & 1915.7 & 1886.4,& 1943.8&-4.219&0.008\\
 2000 & 0.9 & 0.8 & 2   & 1968.6 & 1968.7& 1950.8,& 1985.5&-1.571&0.004\\[1ex] 
 \hline
\end{tabular}
\label{table:4}
\end{table}

The results presented in Table \ref{table:4}
from these simulations show, that when the lists are independent (OR = 1), and the independence model is fitted, the mean of the estimated population size across 2000 multinomial samples is very close to the true population size. However, the bias appears to depend on the size of the population and the response probabilities for each list. When the total population is smaller, N = 1000, and the response probabilities are lower, $\pi_A$ = 0.7 and $\pi_B$ = 0.8, for the lists, the mean of the estimated population sizes is slightly larger than the true population size. 
However, the bias is negligible when N = 1000 and $\pi_A$ = 0.8 and $\pi_B$ = 0.9, as the mean estimated population total across all multinomial samples is approximately equal to the true population total. Therefore, the higher the response probabilities of the lists, the less biased the estimates of the population size will be.

When the lists are dependent, and the independence model is fitted, the mean estimated population size is smaller than the true population size. This shows that positive odds ratios result in population size estimates being negatively biased (compare with Gerritse \textit{et al}., 2015). In some cases, when the odds ratio = 1.1, the true population size falls within the confidence interval. This is an issue of power, as the misspecification of the model is small, and the number of simulations is apparently not large enough to detect it.

\subsubsection{Three-way table}

In our simulation, the three lists have response probabilities 0.8 for list A ($\pi_A$), 0.7 for list B ($\pi_B$) and 0.9 for list C ($\pi_C$). The joint probabilities ($\pi_{111}, \pi_{110}, \pi_{101}, \pi_{100}$, $\pi_{011}, \pi_{010}, \pi_{001}, \pi_{000}$) are, respectively, (0.504, 0.056, 0.216, 0.024, 0.126, 0.054, 0.014 and 0.006). Cell counts are obtained by $\mu_{ijk}=(N\pi_{ijk})$.

\begin{table}[!htb]\footnotesize
\centering
\captionsetup{width=5.6in}
\caption{\small Simulation results for triple systems estimation with inclusion probabilities $\pi_A$ = 0.8, $\pi_B$ = 0.7 and $\pi_C$ = 0.9, with different sizes of odds ratios (OR), specified models, the mean and median of the estimates, the 95\% confidence interval of the mean, the relative bias, and coefficient of variation (CV)}
\begin{tabular}{cllrrrrrr*{1}{S[table-format=2]}} 
 \hline
 N & OR & Model & \multicolumn{1}{c}{Mean} & \multicolumn{1}{c}{Median} & \multicolumn{2}{c}{95\%CI} & \multicolumn{1}{c}{rbias\%} & \multicolumn{1}{c}{CV}\\
\multirow{-2.5}{*} & (AB,AC,BC) & & & & & & &\\
 \hline
 1000 & (1, 1, 1) & [A][B][C] & 1000.0 & 1000.2 & 994.7 & 1004.6 & 0.002 & 0.003\\ 
 1000 & (1, 1, 1) & [AB][AC][BC] & 1000.2 & 1000.1 & 993.5 & 1007.6 & 0.020 & 0.004\\ 
 2000 & (1, 1, 1) & [A][B][C] & 2000.1 & 2000.1 & 1992.9 & 2007.1 & 0.003 & 0.002\\
 2000 & (1, 1, 1) & [AB][AC][BC] & 2000.2 & 2000.1 & 1991.3 & 2010.6 & 0.012 & 0.002\\
 \hline
 1000 & (1.5, 2, 1) & [A][B][C] & 993.7 & 993.9 & 986.7 & 1000.1 & -0.627 & 0.003\\
 1000 & (1.5, 2, 1) & [AB][AC][BC] & 1000.4 & 1000.0 & 990.0 & 1013.0 & 0.036 & 0.006\\
 1000 & (1.5, 2, 1) & [AB][AC] & 1000.2 & 1000.0 & 991.2 & 1009.4 & 0.018 & 0.005\\
 1000 & (1.5, 2, 1) & [A][BC] & 993.0 & 993.2 & 986.0 & 999.4 & -0.704 & 0.004\\
 \hline
 2000 & (1.5, 2, 1) & [A][B][C] & 1987.4 & 1987.4 & 1977.7 & 1996.9 & -0.629 & 0.002\\
 2000 & (1.5, 2, 1) & [AB][AC][BC] & 2000.2 & 1999.6 & 1985.8 & 2017.9 & 0.009 & 0.004\\
 2000 & (1.5, 2, 1) & [AB][AC] & 2000.1 & 2000.2 & 1987.6 & 2014.0 & 0.007 & 0.003\\
 2000 & (1.5, 2, 1) & [A][BC] & 1985.9 & 1985.9 & 1976.2 & 1995.5 & -0.706 & 0.003\\[1ex]
 \hline
\end{tabular}
\label{table:6}
\end{table}

The results of the simulation are shown in Table \ref{table:6}. The model used to generate the population data is specified by the odds ratios in the column OR, and the model used to analyse the data in the column Model. Thus there are correctly specified models, such as OR = (1,1,1) and Model is [A][B][C], overparameterised models, such as OR = (1,1,1) and Model is [AB][AC][BC], and misspecified models, such as OR = (1.5, 2, 1) and Model is [A][BC].
When the model is specified correctly the relative bias is close to zero.
The variance of the estimates, decreases as the population size increases, compare, for example, OR = (1,1,1) and Model [A][B][C] with population sizes 1000 and 2000.
When the model is overparameterised, the variance is larger than when the model is correctly specified.

\section{Discussion}
\label{discussion}

We proposed a simple method to generate contingency tables with fixed marginal probabilities and dependence structures described by loglinear models. We discuss the generation of two-way and three-way contingency tables and this can be extended to in a straightforward way.

We specify the loglinear model in terms of Poisson regression with a log link. Using an offset, prespecified interaction parameters can be fitted that can be understood in terms of odds ratios. For two-way tables, odds ratios describe dependence between two variables and conditional odds ratios describe partial dependence between three or more variables.

We think that this method is simpler than previously proposed methods which are discussed in the introduction, and it can be extended to other models that can be presented as loglinear models, such as logistic regression models with categorical predictors and some latent variable models. The structure of our approach to generating data is the same as the standard approach for undertaking multiple system estimation, so it is straightforward to understand how the dependence is introduced. Our approach is less tailored to the generalised estimating equations which have been the main application of other approaches.

In this paper we apply this method to population size estimation for two-way and three-way contingency tables. The results show the impacts of dependence between lists and model (mis)specification.

\section{Appendix}
\subsection{Population size estimation R code for two-way table}
\begin{verbatim}
### description: general application to population size estimation
### two-way table simulations

#response prob list a
p_a = 0.8
#response prob list b
p_b = 0.7
# total population
N = 1000
# no. iterations
iter = 2000
### odds ratio
osn <- 1

### create dataset from contingency table
list_a <- factor(c("in", "missed", "in", "missed"), levels=c("missed","in"))
list_b <- factor(c("in", "missed", "missed", "in"), levels=c("missed","in"))
count <- c((p_a*p_b*N), ((1-p_a)*(1-p_b)*N),
           ((p_a)*(1-p_b)*N),((1-p_a)*(p_b)*N))
### dataset of frequencies
data_1 <- data.frame(list_a, list_b, count)

### create offset variable
data_1$offset <- c(log(osn), log(1), log(1), log(1))

### model estimated counts from independence model and include offset term
fit_pair_mod22 <- glm(formula = count ~ list_a + list_b,
                      family  = poisson(link = "log"),
                      data    = data_1,
                      offset = data_1$offset)
summary(fit_pair_mod22)
fit_pair_mod22$fitted

data_1$est_countoff<- c(fit_pair_mod22$fitted)

### check to see if offset gave or = osn
or_offset<- (fit_pair_mod22$fitted[1]*fit_pair_mod22$fitted[2])/
  (fit_pair_mod22$fitted[3]*fit_pair_mod22$fitted[4])
or_offset

### check the response of the lists again
### reponse should be 0.8
list_a_resp <- (fit_pair_mod22$fitted[1]+fit_pair_mod22$fitted[3])/
  (fit_pair_mod22$fitted[1]+fit_pair_mod22$fitted[2]+
     fit_pair_mod22$fitted[3]+fit_pair_mod22$fitted[4])
list_a_resp

### response should be 0.7
list_b_resp <- (fit_pair_mod22$fitted[1]+fit_pair_mod22$fitted[4])/
  (fit_pair_mod22$fitted[1]+fit_pair_mod22$fitted[2]+
     fit_pair_mod22$fitted[3]+fit_pair_mod22$fitted[4])
list_b_resp

### total size
data_1$total_samp <- sum(data_1$est_countoff)
data_1_final <- cbind(data_1,data_1$total_samp)

### to store estimates
total_est_store_2 <- NULL
total_pop_store <- NULL
set.seed(123)
### create multinomail samples
for (i in 1:iter) {
  data_1_final$mysum<-rmultinom(1, N, data_1_final$est_countoff)
  data_1_final$mysum <- as.numeric(data_1_final$mysum)
  
  ### throw away missing cells from data
  data_1_final3<-data_1_final[!(data_1_final$list_a=="missed"
                                &data_1_final$list_b=="missed"),]
  
  ### model estimated counts from independence model
  fit_pair_mod2 <- glm(formula = mysum ~ list_a + list_b,
                       family  = poisson(link = "log"),
                       data    = data_1_final3)
  summary(fit_pair_mod2)
  fit_pair_mod2$fitted
  
  ### coefficients from model
  matrix_coef <- summary(fit_pair_mod2)$coefficients
  matrix_coef
  
  ### rounded estimated cell counts
  data_1_final3$est_countoff2<- c(round(fit_pair_mod2$fitted, digits = 0))
  
  ### parameter estimates
  matrix_coef <- summary(fit_pair_mod2)$coefficients
  matrix_coef
  
  ### missing from both cell estimate
  est_total<- exp(matrix_coef[1,1])
  est_total
  
  ### total population of missing cell estimate plus known cell counts
  total_pop = sum(data_1_final3$mysum) + est_total
  
  
  ### store estimates in total_est_store
  total_est_store_2 <- c(total_est_store_2,est_total)
  total_pop_store <- c(total_pop_store, total_pop)}

### 2.5% and 97.5% confidence interval of estimates
### mean and median of estimates
quantile(total_est_store_2, c(.025, .975))
mean(total_est_store_2)
median(total_est_store_2)

### total population outputs
quantile(total_pop_store, c(.025, .975))
mean(total_pop_store)
median(total_pop_store)

#standard deviation
sd<-sd(total_pop_store)
sd
# Relative Bias
rbias<- (mean(total_pop_store)-N)/N
# Relative Bias %
rbias*100
#coefficient of variation
cv<- sd/mean(total_pop_store)
cv
\end{verbatim}

\subsection{Population size estimation R code for three-way table}
\begin{verbatim}

### description: general application to population size estimation
### three-way table simulations
### warning: when specifying the conditional OR's, no more than 4 of them
### should be specified differently.

#response prob list a
p_a = 0.8
#response prob list b
p_b = 0.7
#response prob list c
p_c = 0.9
# total population
N = 1000
# no. iterations
iter = 2000

### odds_ratio
#AB|C=0
osn_1a <- 2
#AB|C=1
osn_1b <- 2
#AC|B=0
osn_2a <- 1
#AC|B=1
osn_2b <- 1
#BC|A=0
osn_3a <- 1
#BC|A=1
osn_3b <- 1

### create dataset from contingency table
list_a <- factor(c("in","in","in","in","missed","missed","missed","missed"), 
                 levels=c("missed","in"))
list_b <- factor(c("in","in","missed","missed","in","in","missed","missed"), 
                 levels=c("missed","in"))
list_c <- factor(c("in","missed","in","missed","in","missed","in","missed"), 
                 levels=c("missed","in"))
count <- c((p_a*p_b*p_c*N),((p_a)*(p_b)*(1-p_c)*N),
           ((p_a)*(1-p_b)*(p_c)*N),((p_a)*(1-p_b)*(1-p_c)*N),
           ((1-p_a)*(p_b)*(p_c)*N),((1-p_a)*(p_b)*(1-p_c)*N),
           ((1-p_a)*(1-p_b)*(p_c)*N),((1-p_a)*(1-p_b)*(1-p_c)*N))
### dataset of frequencies
data_2 <- data.frame(list_a, list_b, list_c, count)

### create offset variable
data_2$offset <- c(log(osn_1b) + log(osn_2b) + log(osn_3b), 
                   log(osn_1a), log(osn_2a), log(1), 
                   log(osn_3a), log(1), log(1), log(1))

### model estimated counts from independence model and include offset term
fit_pair_mod3 <- glm(formula = count ~ list_a + list_b + list_c,
                     family  = poisson(link = "log"),
                     data    = data_2,
                     offset = data_2$offset)
summary(fit_pair_mod3)
fit_pair_mod3$fitted

data_2$est_countoff<- c(fit_pair_mod3$fitted)

#AB|C = 0
### check to see if offset gave or = osn_1a
or_offset<- (fit_pair_mod3$fitted[8]*fit_pair_mod3$fitted[2])/
  (fit_pair_mod3$fitted[4]*fit_pair_mod3$fitted[6])
or_offset

#BC|A = 0
### check to see if offset gave or = osn_3a
or_offset2<- (fit_pair_mod3$fitted[8]*fit_pair_mod3$fitted[5])/
  (fit_pair_mod3$fitted[6]*fit_pair_mod3$fitted[7])
or_offset2

#AC|B = 0
### check to see if offset gave or = osn_2a
or_offset3<- (fit_pair_mod3$fitted[8]*fit_pair_mod3$fitted[3])/
  (fit_pair_mod3$fitted[4]*fit_pair_mod3$fitted[7])
or_offset3

#AB|C = 1
### check to see if offset gave or = osn_1b
or_offset<- (fit_pair_mod3$fitted[1]*fit_pair_mod3$fitted[7])/
  (fit_pair_mod3$fitted[3]*fit_pair_mod3$fitted[5])
or_offset

#BC|A = 1
### check to see if offset gave or = osn_3b
or_offset2<- (fit_pair_mod3$fitted[1]*fit_pair_mod3$fitted[4])/
  (fit_pair_mod3$fitted[2]*fit_pair_mod3$fitted[3])
or_offset2

#AC|B = 1
### check to see if offset gave or = osn_2b
or_offset3<- (fit_pair_mod3$fitted[1]*fit_pair_mod3$fitted[6])/
  (fit_pair_mod3$fitted[5]*fit_pair_mod3$fitted[2])
or_offset3



### check the response of the lists again
### reponse should be 0.8
list_a_resp <- (fit_pair_mod3$fitted[1]+fit_pair_mod3$fitted[2]+
                  fit_pair_mod3$fitted[3]+fit_pair_mod3$fitted[4])/
  (fit_pair_mod3$fitted[1]+fit_pair_mod3$fitted[2]+
     fit_pair_mod3$fitted[3]+fit_pair_mod3$fitted[4]
   +fit_pair_mod3$fitted[5]+fit_pair_mod3$fitted[6]+
     fit_pair_mod3$fitted[7]+fit_pair_mod3$fitted[8])
list_a_resp

### reponse should be 0.7
list_b_resp <- (fit_pair_mod3$fitted[1]+fit_pair_mod3$fitted[2]+
                  fit_pair_mod3$fitted[5]+fit_pair_mod3$fitted[6])/
  (fit_pair_mod3$fitted[1]+fit_pair_mod3$fitted[2]+
     fit_pair_mod3$fitted[3]+fit_pair_mod3$fitted[4]+
     fit_pair_mod3$fitted[5]+fit_pair_mod3$fitted[6]+
     fit_pair_mod3$fitted[7]+fit_pair_mod3$fitted[8])
list_b_resp

### reponse should be 0.9
list_c_resp <- (fit_pair_mod3$fitted[1]+fit_pair_mod3$fitted[3]+
                  fit_pair_mod3$fitted[5]+fit_pair_mod3$fitted[7])/
  (fit_pair_mod3$fitted[1]+fit_pair_mod3$fitted[2]+
     fit_pair_mod3$fitted[3]+fit_pair_mod3$fitted[4]+
     fit_pair_mod3$fitted[5]+fit_pair_mod3$fitted[6]+
     fit_pair_mod3$fitted[7]+fit_pair_mod3$fitted[8])
list_c_resp


### total size
data_2$total_samp <- N
data_2_final <- cbind(data_2,data_2$total_samp)
data_2_final$prob<- data_2_final$est_countoff/data_2_final$total_samp

### to store estimates
total_est_store_2 <- NULL
total_pop_store <- NULL
set.seed(123)
### create multinomail samples
for (i in 1:iter){
  data_2_final$mysum<-rmultinom(1, N, data_2_final$est_countoff)
  data_2_final$mysum <- as.numeric(data_2_final$mysum)
  
  ### throw away missing cells from data
  data_2_final3<-data_2_final[!(data_2_final$list_a=="missed"&
                                  data_2_final$list_b=="missed"&
                                  data_2_final$list_c=="missed"),]
  
  ### model estimated counts from independence model
  fit_pair_mod33 <- glm(formula = mysum ~ list_a + list_b + 
  list_c + list_a*list_b + list_a*list_c + list_b*list_c,
                        family  = poisson(link = "log"),
                        data    = data_2_final3)
  summary(fit_pair_mod33)
  fit_pair_mod33$fitted
  
  ### coefficients from model
  matrix_coef <- summary(fit_pair_mod33)$coefficients
  matrix_coef
  
  ### rounded estimated cell counts
  data_2_final3$est_countoff2<- c(round(fit_pair_mod33$fitted,digits = 0))
  
  ### parameter estimates
  matrix_coef <- summary(fit_pair_mod33)$coefficients
  matrix_coef
  
  ### missing from both cell estimate
  est_total<- exp(matrix_coef[1,1])
  est_total
  
  ### total population of missing cell estimate plus known cell counts
  total_pop = sum(data_2_final3$mysum) + est_total
  
  ### store estimates in total_est_store
  total_est_store_2 <- c(total_est_store_2,est_total)
  total_pop_store <- c(total_pop_store, total_pop)
  
}
### 2.5% and 97.5% confidence interval of estimates
### mean and median of estimates
quantile(total_est_store_2, c(.025, .975))
mean(total_est_store_2)
median(total_est_store_2)

### total population outputs
quantile(total_pop_store, c(.025, .975))
mean(total_pop_store)
median(total_pop_store)

#standard deviation
sd<-sd(total_pop_store)
sd
# Relative Bias
rbias<- (mean(total_pop_store)-N)/N
# Relative Bias %
rbias*100
#coefficient of variation
cv<- sd/mean(total_pop_store)
cv
\end{verbatim}

\section{Disclosure Statement}
The authors report there are no competing interests to declare.

\setstretch{1.0}

\bibliographystyle{agsm}
\bibliography{contingency_tables_fixed_marginal_probabilities_dependence_structures_loglinear_models.bib}

\end{document}